# Rapid aerodynamic prediction of swept wings via physics-embedded transfer learning

Yunjia Yang[*], Runze Li[†], Yufei Zhang[‡]

*Tsinghua University, 100084 Beijing, People's Republic of China*

Lu Lu[§]

*Yale University, New Haven, Connecticut 06511*

and

Haixin Chen[**]

*Tsinghua University, Beijing 100084, People's Republic of China*

**Machine learning-based models provide a promising way to rapidly acquire transonic swept wing flow fields but suffer from large computational costs in establishing training datasets. Here, we propose a physics-embedded transfer learning framework to efficiently train the model by leveraging the idea that a three-dimensional flow field around wings can be analyzed with two-dimensional flow fields around cross-sectional airfoils. An airfoil aerodynamics prediction model is pretrained with airfoil samples. Then, an airfoil-to-wing transfer model is fine-tuned with a few wing samples to predict three-dimensional flow fields based on two-dimensional results on each spanwise cross section. Sweep theory is embedded when determining the corresponding airfoil geometry and operating conditions, and to obtain the sectional airfoil lift coefficient, which is one of the operating conditions, the low-fidelity vortex**

---

[*] Ph.D. Candidate, School of Aerospace Engineering; also visiting assistant in research, Department of Statistics and Data Science, Yale University; yangyj20@mails.tsinghua.edu.cn.

[†] Postdoctoral Research Assistant, School of Aerospace Engineering; lirunze@tsinghua.edu.cn.

[‡] Associate Professor, School of Aerospace Engineering; zhangyufei@tsinghua.edu.cn. Senior Member AIAA.

[§] Assistant Professor, Department of Statistics and Data Science; lu.lu@yale.edu.

[**] Professor, School of Aerospace Engineering; chenhaixin@tsinghua.edu.cn. Associate Fellow AIAA. (Corresponding Author)

**lattice method and data-driven methods are proposed and evaluated. Compared to a nontransfer model, introducing the pretrained model reduces the error by 30%, while introducing sweep theory further reduces the error by 9%. When reducing the dataset size, less than half of the wing training samples are need to reach the same error level as the nontransfer framework, which makes establishing the model much easier.**

# Nomenclature

| | | |
|---|---|---|
| $b$ | = | Span of the wing |
| $c$ | = | Chord length |
| $C_L$ | = | Lift coefficient |
| $C_D$ | = | Drag coefficient |
| $C_{f,\text{s.w.}}$ | = | Streamwise surface friction coefficient |
| $C_{f,z}$ | = | $z$-direction surface friction coefficient |
| $C_p$ | = | Pressure coefficient |
| $(t/c)_{\max}$ | = | Maximum relative thickness |
| $r_t$ | = | Tip-to-root thickness ratio |
| $\alpha$ | = | Angle of attack |
| $\alpha_{\text{twist}}$ | = | Tip-to-root twist angle of the wing |
| $\Lambda_{\text{LE}}$ | = | Leading edge sweep angle of the wing |
| $\Lambda_{1/4}$ | = | 1/4 chord line edge sweep angle of the wing |
| $\Gamma_{\text{LE}}$ | = | Leading edge dihedral angle of the wing |
| $\eta$ | = | Spanwise location of the sectional airfoil |

Subscript:

| | | |
|---|---|---|
| $\infty$ | = | Freestream condition |
| ref | = | Reference operating conditions |

# I. Introduction

Transonic swept wings based on supercritical airfoils have been applied to most modern transportation aircraft, but how to efficiently design them is still an open problem. Because of complex physical phenomena such as shock waves, cross-flow, and separation that may occur on the wing surface, at least the Reynolds-averaged Navier–Stokes (RANS) methods should be applied to obtain a reliable evaluation of transonic wings [1]. However, the unacceptably high computational cost of three-dimensional (3D) simulations makes optimizing the aerodynamic shape difficult for RANS methods. Although adjoint methods have emerged as a promising solution to efficiently optimizing aerodynamic shapes in recent years, they suffer from poor global searching ability, sometimes requiring conducting optimization from several starting points. [2] On the other hand, studies have shown that multi-point and robust optimization are needed to obtain better overall wing performance. [3][4] They require evaluations of wing aerodynamics under multiple design points, which also slows down the optimization process. Therefore, fast prediction methods for transonic swept wing flow fields are still essential for industrial applications.

Since the last century, many fast aerodynamic prediction methods for wings have been proposed. Some of these methods manage to obtain fast but low-fidelity results by solving simplified control equations. The lifting-line method [5][6] and the vortex lattice method (VLM) [7] can provide very fast estimations of aerodynamic coefficients, but they can only be used for thin wings with attached flow. Methods based on potential flow [6] are also widely applied for transonic aerodynamic shape optimization, including the BLWF code [8], which uses an iterative quasi-simultaneous algorithm to solve the strong viscous–inviscid interaction of the external potential flow and boundary layer on lifting surfaces. These methods can quickly estimate the wing surface pressure distribution but cannot precisely predict the shock wave pattern and separation, which are crucial for evaluating the aerodynamic performance of transonic wings [9]. Thus, they can only be used in the preliminary design stages.

Another method for bypassing the time-consuming 3D RANS simulation is to analyze wing performance with a two-dimensional (2D) simulation for sectional airfoils. For transonic swept wings, quasi-2D flow dominates most of the flow around the wing, so 2D sectional airfoil flow fields can act as fair baselines for real 3D flow fields at corresponding spanwise locations along the wing [10]. The difference between 2D and 3D flow fields is known as the 3D effect [11]. To capture this phenomenon, sweep theory is commonly used to correct the 2D flow field considering the sweep angle [11]. Lock et al. [12] and Zhao et al. [13] proposed the 2.75D theory to further consider the taper ratio in the correction. However, the 3D effect near the wing tip and root, as well as under a large angle of attack (AOA), is still difficult to correct.

With the rapid development of machine learning (ML) techniques, especially various deep neural networks, ML-based methods have shown the ability to construct fast and accurate surrogate models for predicting aerodynamic performance and flow fields [14]-[16]. The idea is to first train the model on a relatively large database and then apply it to different aerodynamic optimization problems, so the computational cost investigated in building the database can be shared and finally reduce the evaluation cost [17]. This method provides a new data-driven path for the fast prediction of transonic swept wing flow fields, but meanwhile raises requirements for the generalization ability of the machine-learning model, that is, to train a general model with as few training samples as possible.

In recent years, several frameworks have been tested recently, including dynamic mode decomposition (DMD) [18], multilayer perceptron (MLP) [19]-[22], convolutional neural networks (CNNs) [23][24], Fourier neural operators (FNOs) [25], and graph neural networks (GNNs) [26]. However, the wing geometry considered in most of these studies is limited for simplicity, which prevents them from becoming a general model. The major barrier lies in the large computational burden of establishing a dataset to train the model. A wing needs more parameters to describe its geometry, which means a larger dataset size is needed to cover the space, and the 3D RANS simulation for wings is also time-consuming.

Considering the strong relation between the flow field of a wing and its 2D sectional airfoils, it is natural to combine the traditional 2D to 3D correction with the up-to-date ML methods. The ML-based models can learn the remaining 3D effect other than sweep theory, making the traditional method more accurate. On the other hand, the ML-based model can also benefit from 2D sectional flow fields, as they can act as a prior for 3D flow fields that need to be predicted, so that the complexity of the mapping relationship to be learned is reduced, and thereby the need for 3D wing training samples can be reduced.

Based on this idea, a new ML-based prediction framework for wing aerodynamics is proposed. The key point of the proposed framework is a physics-embedded transfer learning strategy from the 2D airfoil flow field prediction model to the 3D wing flow field prediction tasks. Since airfoil flow field simulation is much quicker than wing flow field simulation, an ML model is first pretrained for predicting airfoil aerodynamics. Then, this model is transferred to the wing by fine-tuning an extra ResNet-based U-Net after it has a few wing samples. During this process, the sweep theory is embedded, so the data-driven model needs to learn only the remaining 3D effect.

The proposed framework shares a similar starting point as the work by Li et al. [22], yet the present work differs from it in several aspects to make the model more practical for applications. First, their work is based on potential flow results that are not precise enough for many downstream tasks, while the present paper uses the RANS simulation results. Second, when predicting the flow field of a new wing, their model requires the spanwise lift distribution (SLD) of the wing to be prescribed, yet in most cases, the SLD is unknown in advance. In the present framework, several solutions, including utilizing VLM and an auxiliary machine-learning model, are proposed and tested to estimate the SLD. Third, the present paper uses a ResNet-based model to transfer the airfoil solutions to the wing flow field for all cross-sections together, instead of a simple MLP network shared for every spanwise location. It improves the model's ability to fit the 3D effect and thereby helps it achieve better accuracy.

## II. Physics-embedded transfer learning framework

### A. Swept wing and sweep theory

The swept wing is used for most modern commercial aircraft. As shown in Fig. 1, a typical swept wing is built by extruding the surface in the $z$-direction between several control sectional airfoils, where each sectional airfoil is in the $x$–$y$ plane. In the present paper, the study object is a single-segment swept wing, where two control sectional airfoils are located at the root and tip.

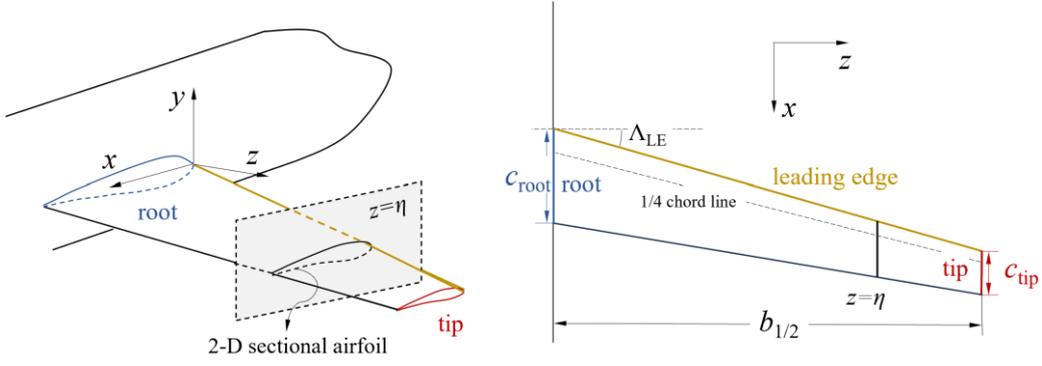

**Fig. 1    Schematic of a single-segment swept wing**

Although the flow field around a wing is 3D, it has a strong relationship with the flow field of its sectional airfoils. At the spanwise location $\eta$, the cross-sectional flow field in the $x$–$y$ plane is highly related to the 2D flow field around the sectional airfoil. When changing the sectional airfoil geometry or the local operating conditions (for example, changing the twist angle of the wing will change the local perceived AOA), the evolution of the 3D flow field will be similar to how the 2D flow field changes. This similarity is known as the local effect, and the difference between the 2D and 3D flow fields is attributed to the 3D effects.

Sweep theory depicts the 3D effects caused by the sweep angle, which contributes a large part of the total effect. This approach provides a fast way to predict the surface pressure distribution. Suppose the wing is under an operating condition of ($\mathrm{Ma}_\infty^{3D}, \mathrm{Re}_\infty^{3D}$), where the Reynolds number is defined with chord $c$. At spanwise location $\eta$, the geometry of the chordwise sectional airfoil is described as $y_\eta^{3D}(x/c_\eta)$, where $c_\eta$ is the local chord. An effective 2D airfoil whose geometry is converted with the sweep angle can be established as follows:

$$y^{2D}(x) = y_\eta^{3D}(x/c_\eta)/\cos\Lambda_{1/4} \qquad (1)$$

and its effective operating conditions are also converted as follows:

$$\begin{aligned}
\mathrm{Ma}_\infty^{2D} &= \mathrm{Ma}_\infty^{3D}\cos\Lambda_{1/4} \\
\mathrm{Re}_\infty^{2D} &= \mathrm{Re}_\infty^{3D}\left(\frac{c_\eta}{c}\right)\cos\Lambda_{1/4} \\
C_L^{2D} &= C_{L,\eta}^{3D}/\cos^2\Lambda_{1/4}
\end{aligned} \qquad (2)$$

where $C_{L,\eta}^{3D}$ is the sectional lift coefficient at spanwise location $\eta$.

If the pressure distribution of the effective 2D airfoil is denoted as $C_p^{2D}(x)$, then the converted sectional pressure distribution at spanwise location $\eta$ of the wing can be obtained with:

$$C_{p,\eta}^{2D \to 3D}\left(x/c_\eta\right) = C_p^{2D}(x) \cdot \cos^2 \Lambda_{1/4} \tag{3}$$

The sweep theory guarantees that the converted sectional pressure distribution is the same as the real 3D results under the assumptions of an infinite span wing and negligible viscosity effects. For an arbitrary swept wing, the 3D effect is more complex because of the cross-flow, and the friction distribution on the wing surface cannot be predicted with sweep theory. However, the sweep theory can still capture a large portion of the effect caused by the sweep angle. This capability inspires us to embed sweep theory into the transfer between the flow field prediction model for airfoils and wings. The data-driven model then needs to learn only the remaining 3D effects, which will be easier.

## B. Transfer learning based on sweep theory

Transfer learning involves a group of techniques for reapplying a model pretrained on one problem (i.e., the source domain) to new problems (i.e., the target domain)[27]. Model-based transfer learning is one of the most common techniques, where the model is first pretrained with a large dataset on the source domain, and then, the model or part of the model can be fine-tuned with only a few samples on the target domain to obtain good accuracy.

For the proposed transfer learning framework, the source and target domains are the 2D airfoil and 3D wing, respectively. As shown in Fig. 2, the transfer learning framework is designed to pretrain a model on an airfoil dataset and fine-tune it with a few wing samples. The difference between the 2D and 3D values is modeled with sweep theory and extra neural network layers to capture the remaining effects. The pretrained and fine-tuned parts are called the airfoil aerodynamics model and the airfoil-to-wing transfer model, respectively.

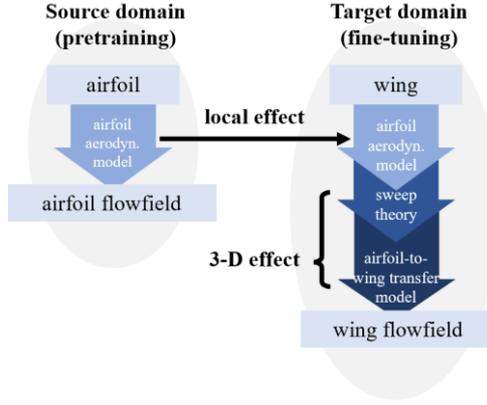

**Fig. 2   Source and target domains of the transfer learning framework**

The flowchart of the proposed transfer learning framework is shown in Fig. 3, where the input and output of each model are shown in gray boxes. The abbreviation S.T. in the figure indicates that the corresponding sweep theory equation is applied to transform the values. In the following sections, the pretraining and fine-tuning stages are introduced.

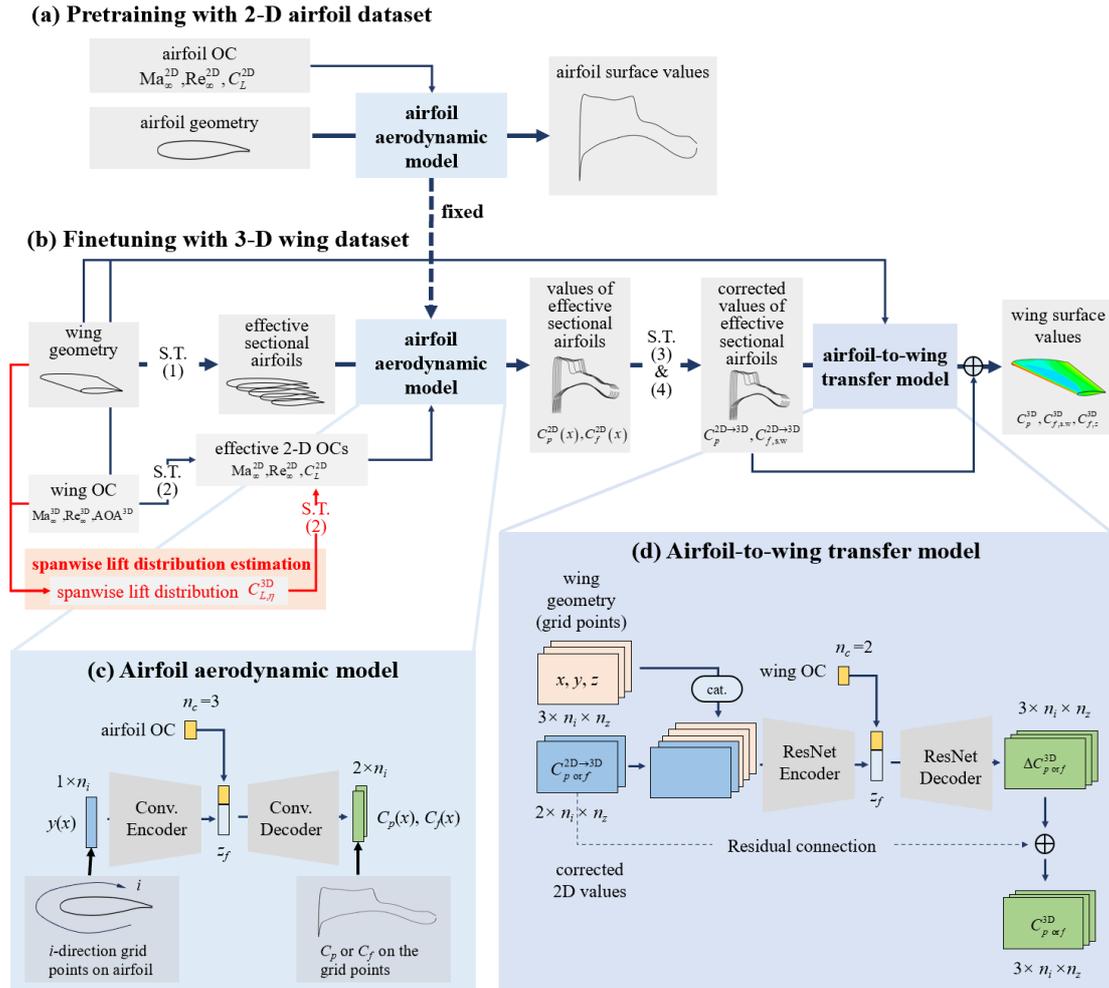

**Fig. 3    Overall framework of physics-embedded transfer learning**

## 1.    Pretraining of the airfoil aerodynamics model

During pretraining, the airfoil aerodynamics model is trained with airfoil samples. As shown in Fig. 3 (c), the input of the model is the airfoil geometry and operating conditions, which include the Mach number, the Reynolds number, and the lift coefficient. The airfoil is described with the $y$ coordinates of the $n_i$ airfoil circumferential direction ($i$-direction) grid points on the airfoil surface. The model outputs the surface pressure and friction coefficient distributions on the airfoil's surface at the same grid points as the input geometry. Since the input and output are one-dimensional (1D), the model is realized with a U-Net based on 1D convolutional neural networks, and its detailed architecture can be found in Appendix A1.

## 2.    Fine-tuning of the airfoil-to-wing transfer model

After the airfoil aerodynamics model is well trained, it is transferred to predict the 3D flow field of wings by fine-tuning with wing samples. For each wing sample, $n_z$ spanwise locations are evenly selected from the root to the tip. The cross-sectional airfoil geometries and local operating conditions at these spanwise locations are transformed via sweep theory Equations (1) and (2) to obtain the effective 2D airfoil geometries and operating conditions. An issue emerges here, as the sectional lift coefficient of the wing $C_{L,\eta}^{3D}$ in Equation (2), i.e., the spanwise lift distribution (SLD), is unknown when predicting new wing samples. In the following section, several methods are proposed for estimating the SLDs for the training and predicting stages.

With the effective 2D airfoil geometry and operating conditions, the pressure and friction distribution, denoted as $C_p^{2D}(x), C_f^{2D}(x)$, can be obtained with the pretrained airfoil aerodynamics model. Then, sweep theory is used again to transform the 2D distributions to corrected values with Equation (3). Here, the surface friction distribution coefficients are also transformed from 2D to 3D with a formula that imitates (3):

$$C_{f,s.w.,\eta}^{2D \to 3D}\left(x/c_\eta\right) = C_f^{2D}(x) \cdot \cos^2 \Lambda_{1/4} \qquad (4)$$

where the subscript s.w. represents the streamwise component of the surface friction coefficient on the wing surface, as shown in Fig. 4.

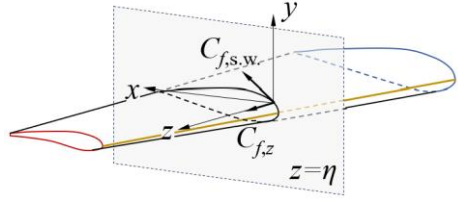

**Fig. 4    Schematic for the decomposition of the surface friction coefficient**

The corrected 2D values and the real 3D values are used to train the airfoil-to-wing transfer model for the fine-tuning stage. As shown in Fig. 3 (d), the model input contains three parts: the corrected pressure and friction distributions predicted with the airfoil aerodynamics model, the wing geometry, and the operating conditions ($Ma_{inf}$ and AOA). For the model's output, the corrected 2D values are subtracted from the pressure and streamwise coefficient distributions of the real 3D surface distributions so that the model needs to fit only the difference between the corrected 2D and 3D surface distributions.

The corrected distributions and the final output are two-dimensional data. The first dimension represents the *i*-direction grid points on each cross-sectional airfoil, which is the same as in the airfoil aerodynamics model, and the second dimension represents the spanwise location. The input values have two channels corresponding to the corrected surface pressure and streamwise friction distribution, and the output values have one more channel for the *z*-direction friction distribution. The U-Net based on 2D ResNet is used as the backbone of the airfoil-to-wing transfer model, whose detailed architecture is described in Appendix A2.

## 3.    Spanwise lift distribution estimation

In the overall framework of physics-embedded transfer learning, the SLD is used as an input for sweep theory to determine the corresponding effective operating conditions for a 2D sectional airfoil. The SLD must be obtained before the aerodynamics of the wing are predicted, so additional efforts are needed to predict the SLD directly from the wing geometry and operating conditions.

Since SLD is important for wing aerodynamic and aeroelastic analysis, several low-fidelity semi-analytical methods are developed for estimating it. The strip theory [28] and the lifting line theory [5] can analytically predict the lift curve slope at different spanwise locations, and along with the zero-lift angle of attack predicted with the thin-airfoil theory, the SLD can be estimated. The vortex lattice method (VLM) [7] can give a little more precise estimation by solving the lifting surface problem. The wing surface is discretized into lattices, and a horseshoe vortex is assumed to be generated by every lattice that will induce velocity. By solving the normal velocity equals zero for every lattice, the pressure distribution can be obtained, and thereby, the SLD.

These methods can provide fair SLD results for wings with thin airfoils and low speeds but have a large error when the flow over the wing is more complex, such as for a transonic swept wing, since they are based on the potential flow [7]. However, these results can still provide a reasonable reflection of how the wing geometry and operating conditions affect the SLD.

Utilizing the above-mentioned VLM algorithm, three methods for estimating SLDs are proposed and compared in the present paper: (1) **Low-fidelity estimation**: using the SLD calculated with the low-fidelity VLM algorithm; (2) **Data-driven estimation**: a purely data-driven method is used to train a model for SLD prediction; (3) **Combination estimation**: The ML model is used to predict the residual between the VLM results and the real SLD.

The open-source code PyTornado [29] is used in the present study for VLM calculations, and for the data-driven part, an auxiliary ML model based on feed-forward neural networks (FNNs) is built to predict the lift distribution or the residual. The detailed architecture for the ML model can be found in Appendix A3.

## III. Dataset establishment

Two datasets are established in the present study to train the airfoil aerodynamics model and the airfoil-to-wing transfer model. The airfoil model is trained with a supercritical airfoil dataset, while the transfer models, including the lift distribution model and the airfoil-to-wing model, are trained with a transonic swept wing dataset.

### A. Sampling of airfoil and wing geometries

# 1. Geometry parameterization of the swept wing

Several parameters are used to describe the geometry of a swept wing, and the first part describes its control sectional airfoils. Their geometry is parameterized by the class shape transformation (CST) method with ninth-order Bernstein polynomial basis functions. The upper and lower surfaces of the airfoil are described separately, which leads to 20 independent variables denoted as $(u_i, l_i)$ $i = 0,1,\ldots 9$. The maximum relative thickness $(t/c)_{max}$ and chord length $c$ also need to be prescribed when reconstructing airfoil geometry with CST parameters.

For simplicity, the root airfoil is set to have a chord of 1 on the $x$-axis with its leading edge located at the origin for all wings. The tip airfoil has the same CST parameters as the root airfoil, but its maximum thickness and chord length can be different. The ratios between the maximum tip and root thickness and between the tip and root chord length are called the thickness ratio $r_t$ and the taper ratio TR, respectively. Then, other planform geometry parameters, including the sweep angle ($\Lambda_{LE}$), dihedral angle ($\Gamma_{LE}$), aspect ratio (AR), and twist angle ($\alpha_{twist}$), are used to determine the position and rotation of the tip airfoil, thereby describing the shape of the wing. As shown in Fig. 5, the sweep and dihedral angles are the angles between the $z$-axis and the projection of the leading edge on the $x$–$z$ plane and the $y$–$z$ plane, respectively. The aspect ratio is defined as $AR = b_{1/2}^2 / 2S_{1/2}$, where $b_{1/2}$ is the half span and $S_{1/2}$ is the wing projection area. The twist angle is defined as the rotation angle of the tip airfoil to the root airfoil according to the $z$-axis, and its sign is defined with the right-hand rule on the $z$-axis.

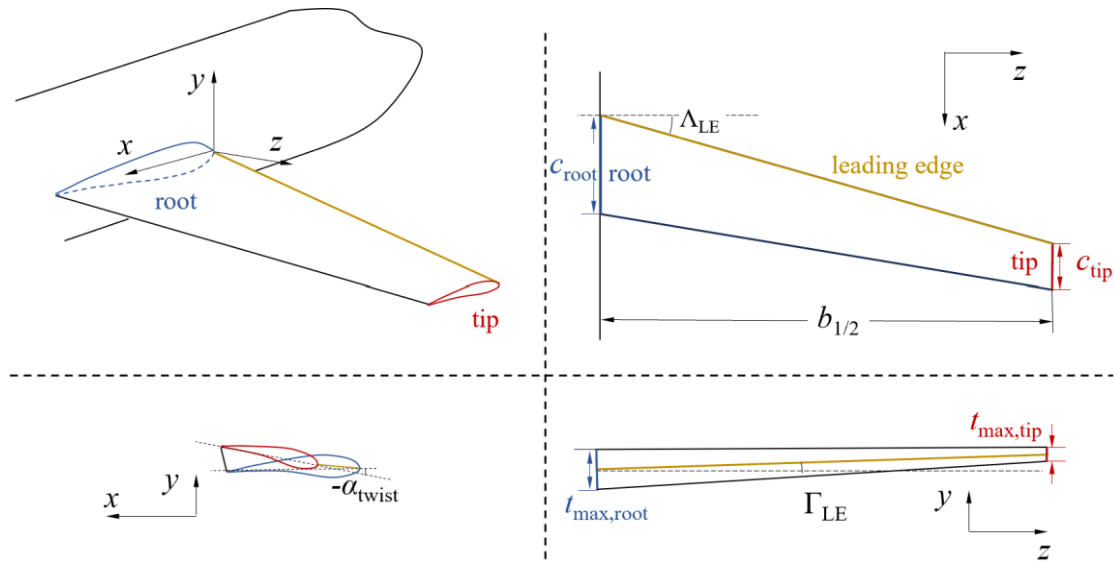

**Fig. 5    Definition of the geometric parameters for a single-segment wing**

## 2.  Sampling of airfoil geometry

The supercritical airfoil dataset in this paper was built according to airfoil geometries in our previous study[17], where the output space sampling (OSS) method[30] was used to sample the geometric parameters. The OSS is an adaptive sampling method that aims to obtain airfoils with more abundant and diverse pressure distribution patterns under reference operating conditions.

First, several reference operating conditions, namely, the free stream Mach number ($Ma_{\infty,ref}$), the airfoil lift coefficient ($C_{L,ref}$), and the freestream Reynolds number, are assigned. The lift coefficient $C_L$ is used instead of AOA in airfoil sampling because its range can be more easily determined with the sweep theory. During CFD simulation, the algorithm automatically adjusts the AOA of the airfoil to fulfill the assigned $C_L$. Moreover, several maximum relative thicknesses are also selected. Then, the OSS is used to sample the airfoil CST parameters to identify abundant pressure distribution patterns. The sampling ranges of the parameters in this study are listed in Table 1, and 1420 groups of CST parameters are obtained. Several airfoil samples are plotted in Fig. 6, with the maximum thickness increasing from left to right.

**Table 1.    Sampling ranges of airfoil parameters in the output space sampling**

| parameters | description | upper boundary | lower boundary |
|---|---|---|---|
| $u_0, u_1, \ldots, u_9$ | upper surface CST parameters | -2.0 | 2.0 |
| $l_0, l_1, \ldots, l_9$ | lower surface CST parameters | -2.0 | 2.0 |
| $(t/c)_{max}$ | maximum relative thickness | 0.09 | 0.13 |
| $Ma_{\infty,ref}$ | reference freestream Mach number | 0.71 | 0.76 |
| $C_{L,ref}$ | reference lift coefficient | 0.60 | 0.90 |

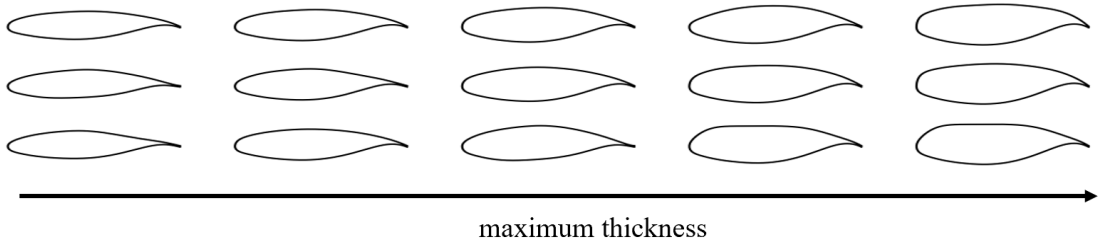

maximum thickness

**Fig. 6    Examples of airfoils**

### 3. Sampling of the swept wing geometry

The swept wing geometries are built from the supercritical airfoil dataset. For each group of CST parameter samples in the supercritical airfoil dataset, two single-segment wings are generated. The remaining wing geometric parameter values, including the root airfoil maximum thickness and planform parameters, are generated via random sampling over a wide range of swept wings at different speeds. The ranges for the parameters are listed in Table 2. A total of 2800 wing geometries are obtained in the process. The geometric projections on the $x$–$z$ plane of several wing samples with different sweep angles, taper ratios, and aspect ratios are plotted in Fig. 7 to show the various wing geometries in the dataset.

Table 2. Sampling range for the wing geometric parameters

| Description | Parameter | Upper bound | Lower bound |
|---|---|---|---|
| baseline airfoil maximum relative thickness | $(t/c)_{max,root}$ | 0.09 | 0.13 |
| thickness ratio | $r_t$ | 0.8 | 1.0 |
| sweep angle | $\Lambda_{LE}$ | 0° | 35° |
| dihedral angle | $\Gamma_{LE}$ | 0° | 3° |
| aspect ratio | AR | 6 | 10 |
| taper ratio | TR | 0.2 | 1.0 |
| twist angle | $\alpha_{twist}$ | 0° | -6° |

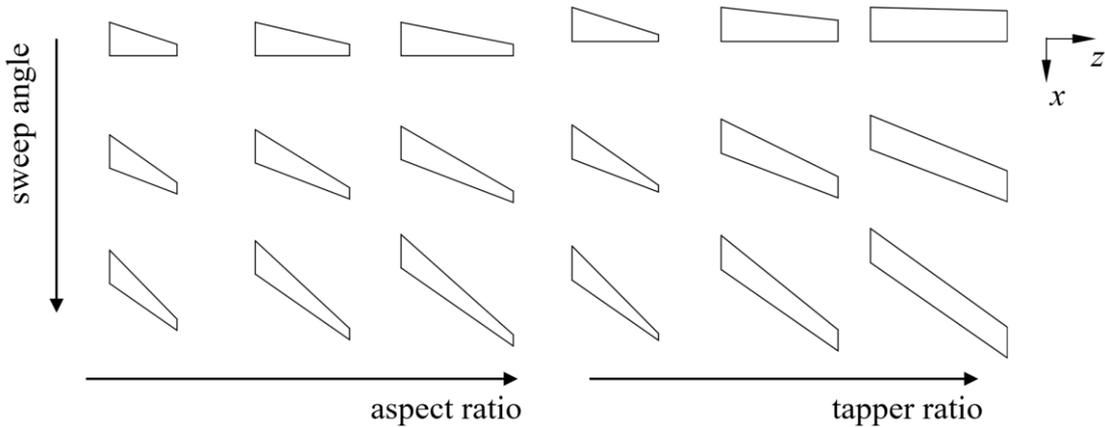

Fig. 7   Examples of wing projection on the $x$–$z$ plane

### B. Sampling of airfoil and wing operating conditions

The operating conditions of the wings include the freestream Mach number ($Ma_\infty$), Reynolds number based on unit length (Re), freestream temperature $T_\infty$, and angle of attack (AOA). In the present dataset, one operating condition is assigned to each wing geometry, and the ranges for the conditions are listed in Table 3.

Table 3.  Sampling ranges of the wing operating conditions

| parameters | description | upper boundary | lower boundary |
|---|---|---|---|
| AOA | angle of attack | 1° | 6° |
| $Ma_\infty$ | freestream Mach number | 0.72 | 0.85 |
| $Re_\infty$ | freestream Reynolds number based on unit length | 6.43 million | |
| $T_\infty$ | freestream temperature | 580°R | |

For the airfoils, 24 operating conditions are selected separately for each airfoil geometry according to the operating condition of the corresponding wing that uses the airfoil as the baseline. The operating conditions of the airfoil are described by $Ma_\infty$, $Re_\infty$, $T_\infty$, and lift coefficients ($C_{LS}$), which are randomly sampled from the ranges that cover the corresponding values of the wing via sweep theory. Suppose that the wing has a sweep angle of $\Lambda$; then, the sampling ranges for the airfoil operating conditions can be found in Table 4, where $C_{L,\eta}^{3D}$ is the sectional lift coefficient along the spanwise direction of the corresponding wing.

Table 4.  Sampling ranges of the airfoil operating conditions

| parameters | description | upper boundary | lower boundary |
|---|---|---|---|
| $C_L$ | lift coefficient | $\left(\min_\eta C_{L,\eta}^{3D}\right)/\cos^2\Lambda$ | $\left(\max_\eta C_{L,\eta}^{3D}\right)/\cos^2\Lambda$ |
| $Ma_\infty$ | freestream Mach number | $Ma_\infty \cdot \cos\Lambda$ | |
| $Re_\infty$ | freestream Reynolds number based on unit length | $Re_\infty^{3D} \cdot TR \cdot \cos\Lambda$ | $Re_\infty^{3D} \cos\Lambda$ |
| $T_\infty$ | freestream temperature | 580°R | |

## C. CFD settings

### 1. Computational grid

The structured grids are used for 2D and 3D simulations and are generated with the in-house code CGrid. The 2D C-type grid is the same as that in Ref.[17]. This grid has 401 points in the airfoil circumferential direction (*i*-direction) and 81 points in the wall-normal direction (*j*-direction). The *i*-direction has 321 points on the airfoil surface, which are distributed more densely near the leading and tailing edges, and 41 points in the wake region. For the wall-normal direction, the height of the first grid layer is set to fulfill $\Delta y^+ < 1$, and the far field is located 20 chords away from the airfoil.

The 3D grid for the wing simulation is the same as that in Ref. [31]. This grid is extruded from 2D section grids in the spanwise direction. The spanwise grid has 61 points, which are distributed more densely near the wing tip. Then, the grid points near the wing tip are extruded outside to form an O-type grid to help improve the grid quality. Finally, the outermost grid section is extruded to the far field in the spanwise direction, which is 30 chords away from the tip, with 41 grid points. The grid size is 2.86 million cells for single wings.

2. **Computational setting**

The airfoil and wing flow fields are computed using the open-source Reynolds-averaged Navier–Stokes (RANS) solver CFL3D[32][33], which has been widely employed in engineering applications. This solver is based on the finite volume method, where the MUSCL scheme, ROE scheme, and implicit Gauss-Seidel algorithm are adopted for flow variable reconstruction, spatial discretization, and time advancement. The shear stress transport (SST) model is adopted for turbulence modeling. A three-level W-cycle multigrid is adopted to accelerate the calculation. The steps for the three multigrid levels are 1000, 1000, and 2000 for 2D airfoil simulations, and 6000 more third-level steps are used to achieve convergence for 3D wings. More detailed CFD settings and the validation of current CFD settings can be found in our previous works[17][31]. After the calculation, flow fields for 22104 airfoils and 1842 wings were obtained, while the calculations of the others were not convergent.

D. **Postprocessing**

The pressure and friction coefficient distributions are extracted from the simulated flow fields for training and testing. The pressure can be directly obtained from every surface grid point on the airfoil or the wing, and the friction at every surface grid point is calculated from the gradient of the tangential velocity between the first and second grid layers. For 2D airfoils, the friction is a scalar whose sign is positive when the tangential velocity has the same direction with increasing grid $i$-index. For 3D wings, the friction is decomposed into a streamwise part $C_{f,\text{s.w.}}$ in the $x$–$y$ plane and a spanwise part $C_{f,z}$ in the $z$-direction, as shown in Fig. 4. The streamwise part has the same sign definition as for the 2D airfoils.

The surface pressure and friction coefficient distributions are then interpolated from the computing surface grid points to an aligned grid for the airfoils and the wings. For the airfoil circumferential direction ($i$-direction), a fixed series of relative $x$-positions $\{(x/c)_i\}$ is used for the 2D airfoils and the sectional values for the 3D wings. Considering the two grids are quite similar, the interpolation on the $x$-direction will not bring much error. In the spanwise direction ($j$-direction), 101 $x$–$y$ sectional planes that are evenly distributed in the span are used. This procedure gives the geometry grid and surface values of 321×1 points for the airfoil surface and 321×101 points for the wing surface. The sectional lift coefficients are also obtained for wing samples during this process by integrating the surface pressure and streamwise friction coefficients on each of the 101 sectional planes.

The surface pressure and friction coefficient distributions, as well as other input and output parameters in the present study, are nondimensionalized with their maximum and minimum values.

# IV. Model training and results

## A. Airfoil aerodynamics prediction

The airfoil aerodynamics model serves as a basic model to be transferred to the framework and is pretrained with the 2D airfoil dataset in advance. In this section, the performance of the pretrained model is evaluated.

During training, the airfoil dataset is split into two parts, where 90% of the samples are used to train the model, and the remaining 10% are used for evaluating model performance. The mean square error (MSE) between the model-predicted and CFD-simulated airfoil surface distributions is used as the loss function. The training settings are similar to those in Ref.[31]: A batch size of 16 is applied, and the optimizer is the Adam algorithm. The learning rate follows a warmup strategy where it is increased from $1 \times 10^{-5}$ to $1 \times 10^{-4}$ in the first 20 epochs and then reduced by an exponential function with a base of 0.95. The training process has 300 epochs, and in each epoch, all mini-batches are shuffled randomly and used iteratively to train the model.

The training process is conducted five times separately with the same training samples but different random initializations of the trainable parameters at the beginning. Among the five runs, the model with the smallest surface coefficient distribution error on the test airfoil samples is used for the subsequent transfer learning. For each test sample, the relative MSE for each coefficient is obtained and divided by the range of the coefficient. Then, the test error is calculated by averaging the relative MSE among the test samples as follows:

$$\delta C_{p \text{ or } f} = \frac{1}{N_s} \sum_{n=0}^{N_s} \frac{\sqrt{\sum_{i=0}^{N_i}\left(C_{p \text{ or } f}^{\text{model}} - C_{p \text{ or } f}^{\text{ground truth}}\right)^2}}{\max_i\left(C_{p \text{ or } f}^{\text{ground truth}}\right) - \min_i\left(C_{p \text{ or } f}^{\text{ground truth}}\right)} \times 100\% \tag{5}$$

Table 5 shows the smallest, average, and standard deviation of the test errors among the five runs. The smallest test errors are below 0.2%. The performance of the model is also illustrated for three airfoil samples with the largest errors. Their geometries, operating conditions, and model-predicted and CFD-simulated surface distributions are depicted in Fig. 8. For samples with the largest errors, the model-predicted distributions can well fit the ground truth, which indicates the excellent performance of the airfoil aerodynamic model and paves the way for subsequent fine-tuning.

**Table 5.    Prediction errors of the airfoil aerodynamics model**

| variables | smallest test error (Run No. 4) | average test error ± standard deviation |
|---|---|---|
| $C_p$ | 0.189% | (0.230 ± 0.035) % |
| $C_f$ | 0.083% | (0.087 ± 0.005) % |

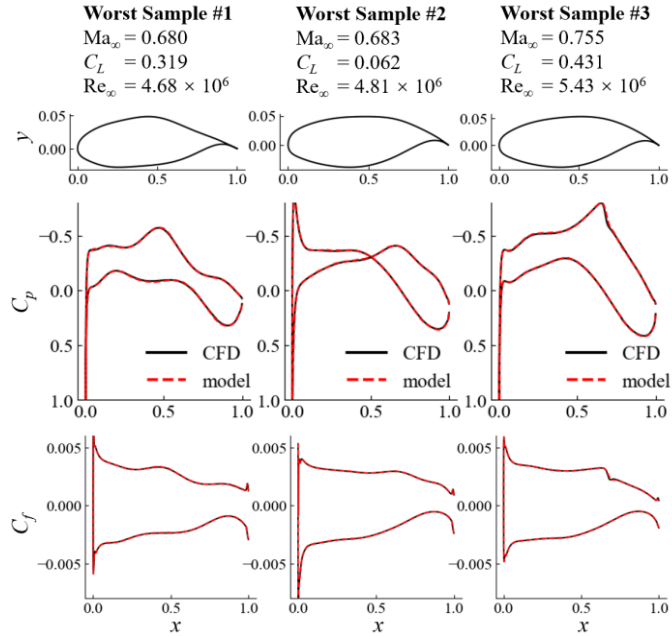

**Fig. 8    Surface distribution predicted by the airfoil aerodynamics model for three samples**

**with the largest error**

## B. Wing aerodynamics prediction with real SLD

After the airfoil aerodynamics model is well-trained, the airfoil-to-wing transfer model is trained on the wing dataset as the fine-tuning stage. In this section, the improvements resulting from the use of the pretrained aerodynamics model and embedded sweep theory are evaluated.

In this section, the SLDs used during the process of obtaining the surface distributions of 2D airfoils are from real 3D simulations for training and testing samples. As mentioned before, the SLDs are unknown before prediction in real applications. We use real SLDs here to evaluate the improvements of introducing transfer learning and sweep theory without being bothered by the error in SLD estimation. This part of the error will be discussed in the following section.

Two baseline frameworks for wing aerodynamics prediction are established, as shown in Fig. 9. The nontransfer framework (Fig. 9 (a)) completely abandons the pretrained model and only uses the wing geometry and operating conditions as inputs for the fine-tuning model. It has the same architecture as the airfoil-to-wing transfer model, except that the input of the 2D ResNet encoder contains only the wing geometry. The direct transfer framework (Fig. 9 (b)) fine-tunes the airfoil-to-wing model without considering sweep theory. The wing sectional airfoil geometry and wing operation conditions ($\text{Ma}_\infty^{3D}$, $\text{Re}_\infty^{3D}$, $C_{L,\eta}^{3D}$), rather than the effective conditions, are input to the airfoil aerodynamics model, and the output 2D airfoil surface distributions are also directly used for the airfoil-to-wing transfer model without being corrected. The physics-embedded transfer framework (Fig. 9 (c)) uses only the model introduced in Section II.

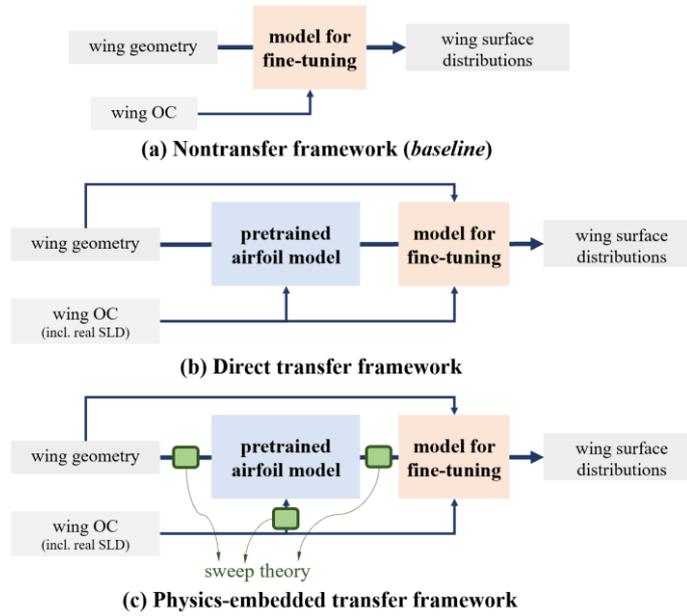

**Fig. 9   Three frameworks for testing improvements in physics-embedded transfer learning**

### 1.   Surface distributions of effective sectional airfoils

Fig. 10 shows the wing surface distribution predicted with the pretrained airfoil aerodynamics model from the effective 2D sectional airfoils. The first column shows the results that do not consider sweep theory when deciding the geometry and operating conditions of effective sectional airfoils (corresponding to the direct transfer framework), and the second column shows the results that consider sweep theory.

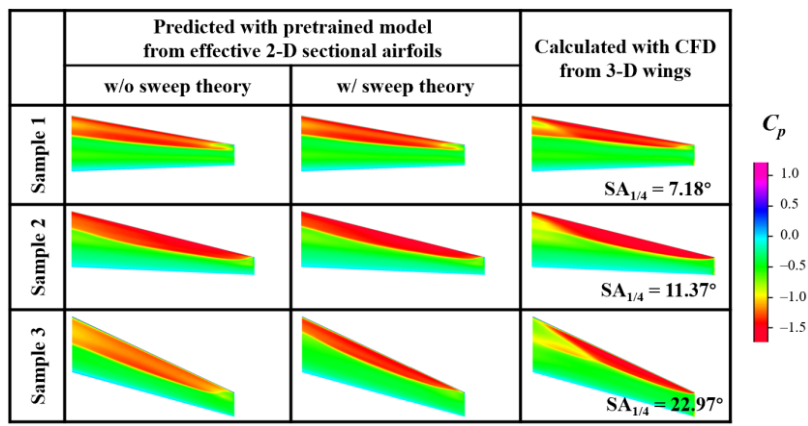

(a) Surface pressure coefficient distributions

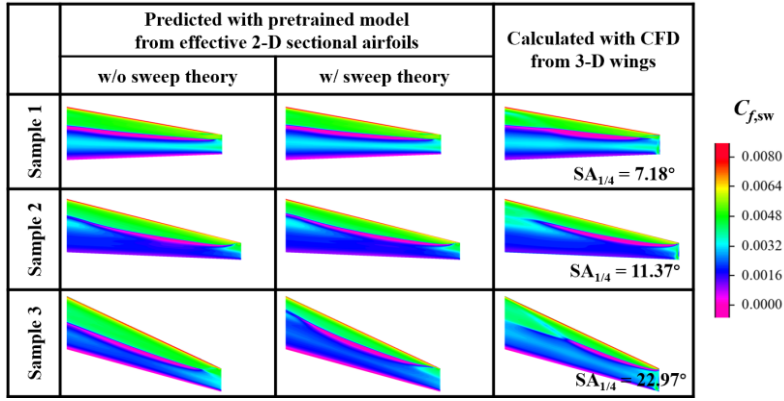

(b) Surface streamwise friction coefficient distributions

Fig. 10 Upper Surface contours of wings predicted by the pretrained airfoil aerodynamics model from their 2D sectional airfoils

For wings with small sweep angles, the surface distributions of the effective sectional airfoils are quite similar to the real 3D calculated distributions regardless of whether sweep theory is considered, and the difference is mostly located near the tip and the root. This comparison indicates that the local effect has a dominant influence on the flow field, especially at small sweep angles. As the sweep angle increases, the difference increases, and the method that considers sweep theory can provide a more reasonable prediction.

## 2. Wing prediction performance comparison of frameworks with and without transfer learning

Using the surface distributions of effective sectional airfoils, the airfoil-to-wing transfer model can be trained. Tenfold cross-validation was used to fine-tune and evaluate the three frameworks involved in the comparison. The wing dataset is first randomly split into 10 groups. Then, the training process was repeated ten times, and each time, the training samples were divided into nine of the ten groups, while the remaining group was used to evaluate model performance. Since the dataset is smaller, the batch size is set to 8, and the learning rate schedule is prolonged three times compared with the training of the airfoil aerodynamics model, resulting in 900 epochs of training. The rest of the training setup remains the same.

The cross-validation scores of the three frameworks are compared in Table 6. The metrics are the relative MSEs for the pressure, streamwise friction, and $z$-direction friction coefficients, which are defined in the same way as Equation (5). The coefficients of lift, drag, and pitching moment according to the 1/4 chord point at the root are obtained by integrating the predicted wing surface distribution and are evaluated with mean absolute errors (MAEs) with the ground truth. The average accuracy gain for each metric and each framework is also presented by using the nontransfer framework as the baseline and calculating the ratio of the error reduction to the baseline error. The accuracy gains are shown in brackets below each error value.

Table 6. Prediction errors of frameworks with and without transfer learning

|  | w/o transfer | w/ direct transfer | w/ physics-embedded transfer |
|---|---|---|---|
| $C_p$ | 1.08% ± 0.06% | 0.75% ± 0.04% (-30.7%) | 0.66% ± 0.04% (-39.3%) |
| $C_{f,\text{sw}}$ | 0.65% ± 0.04% | 0.47% ± 0.03% (-28.0%) | 0.42% ± 0.02% (-35.6%) |
| $C_{f,z}$ | 0.35% ± 0.02% | 0.28% ± 0.02% (-20.6%) | 0.26% ± 0.01% (-24.7%) |
| $C_L \times 10^3$ | 8.08 ± 0.37 | 2.12 ± 0.22 (-73.7%) | 1.70 ± 0.17 (-79.0%) |
| $C_D \times 10^4$ | 5.08 ± 0.41 | 3.35 ± 0.31 (-34.1%) | 2.87 ± 0.31 (-43.5%) |
| $C_{Mz} \times 10^3$ | 4.77 ± 0.43 | 1.44 ± 0.11 (-69.8%) | 1.28 ± 0.14 (-73.1%) |

As mentioned before, a model that directly predicts wing flow fields needs to learn the local effect and 3D effects. The introduction of the pretrained aerodynamics model helps capture the local effect before the fine-tuning stage and contributes a 20% to 30% increase in prediction accuracy for surface distributions. Further introducing sweep theory by physics-embedded transfer learning will help the model capture part of the 3D effect caused by the sweep angle, which contributes another 4% to 9% accuracy gain, indicating that sweep-angle-induced 3D effects are quite important for prediction. The pressure coefficient has the largest increase among the three distributed quantities, which is reasonable because the 3D effect has less influence on surface pressure than does surface friction, and the sweep theory is proposed for transfer surface pressure distributions. The results demonstrate that physics-embedded transfer learning can increase the accuracy of the wing friction distribution.

Among the integral aerodynamics coefficients, the accuracy gains for the drag coefficients are slightly larger than those for the distributed surface quantities, while a dramatically larger gain of approximately 70% is obtained for the lift and pitching moment coefficients. The reason is likely the real

SLD input for the airfoil aerodynamics model. The real SLD input guarantees that the 2D sectional surface distributions predicted with the airfoil aerodynamics model have almost correct lift coefficients at each spanwise location, which means that the input for the airfoil-to-wing transfer model already has almost correct wing lift coefficients. The transfer model will learn the remaining corrections more readily. When using the estimated SLD in real applications, the accuracy gains for the lift and pitching moment coefficients should be smaller, which will be verified in the following section.

To intuitively illustrate the model accuracy, the CFD-simulated and model-predicted surface distributions are depicted and compared. The model used here is the physics-embedded transfer framework with real SLDs. Three samples with the largest pressure distribution prediction errors are selected, and their prediction results are obtained by the model trained on which the sample is used as the test set in ten cross-validation runs. Thus, the results below show the model performance under the test scenario.

Fig. 11 shows the prediction results of several example wings. In each row of subfigure (a), the surface pressure, streamwise friction, and spanwise friction coefficient contours are shown from top to bottom. The model-predicted contours are plotted on the right side, and the CFD-simulated results are plotted flipped on the left side. In subfigure (b), the surface pressure distributions on three cross sections at 20%, 50%, and 80% spanwise locations for each wing are depicted. The CFD-simulated profiles are depicted with black solid lines, while the results of the nontransfer and physics-embedded transfer models are drawn as red and blue dashed lines, respectively.

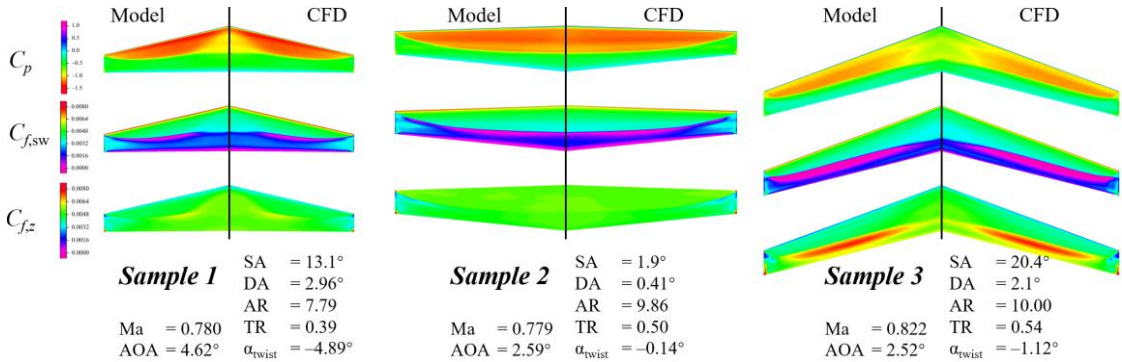

(a) Upper surface contours

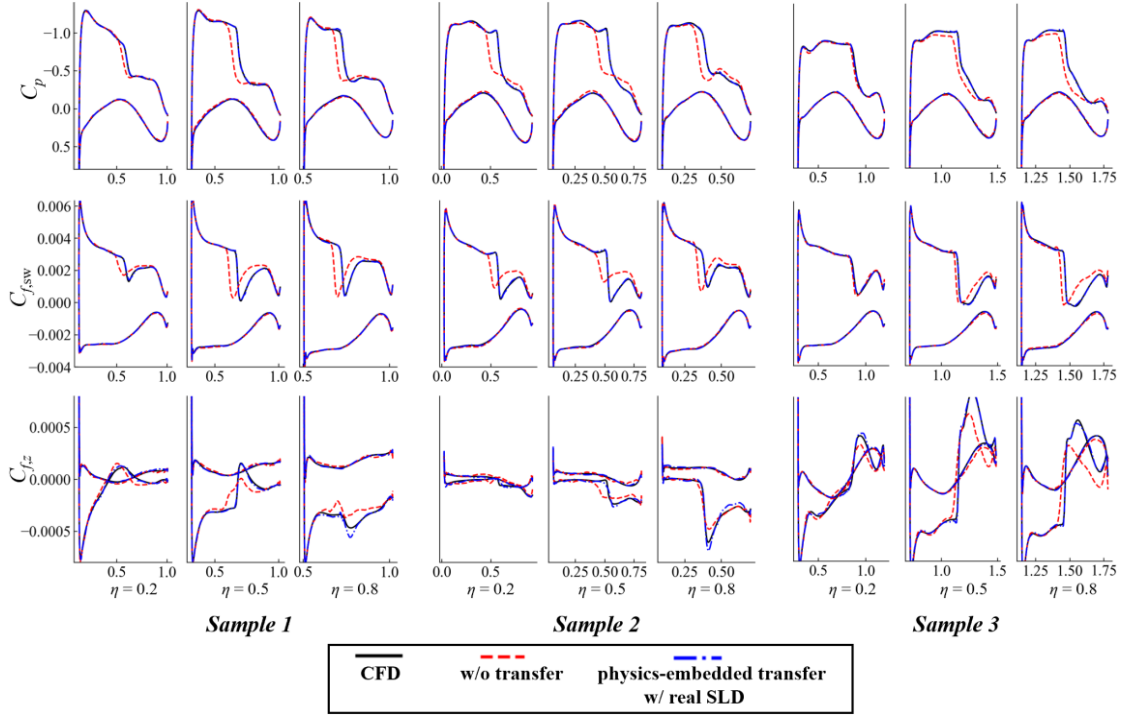

(b) Cross-sectional distributions

**Fig. 11** Results of the example wings predicted with CFD and the transfer model with real SLDs

The results show that the accuracy gain between the frameworks with and without transfer learning is obvious, as the proposed physics-embedded transfer learning framework can capture important flow features such as shock waves and recirculation zones around transonic wings, even in the worst cases.

### 3. Wing prediction performance with different training dataset sizes

The cross-validation scores of the models trained on different numbers of training samples are evaluated and shown in Fig. 12. When the model is not trained on the entire dataset, a given number of samples is randomly selected from the training dataset for each cross-validation run, and the model is trained on selected samples. The testing dataset is still the on-hold group for each run, and the final score is obtained by averaging the scores of each cross-validation run. The mean errors of the ten cross-validation runs are shown as the symbols and lines, and the standard deviations are shown as the error bars.

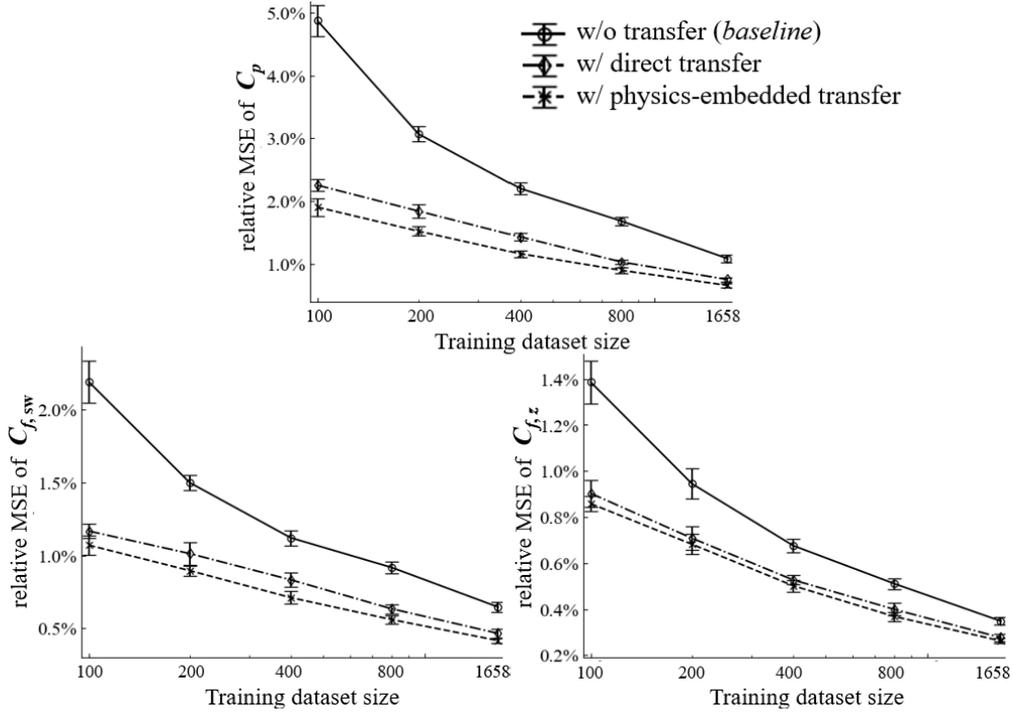

(a) Mean square error of distributed flow quantities

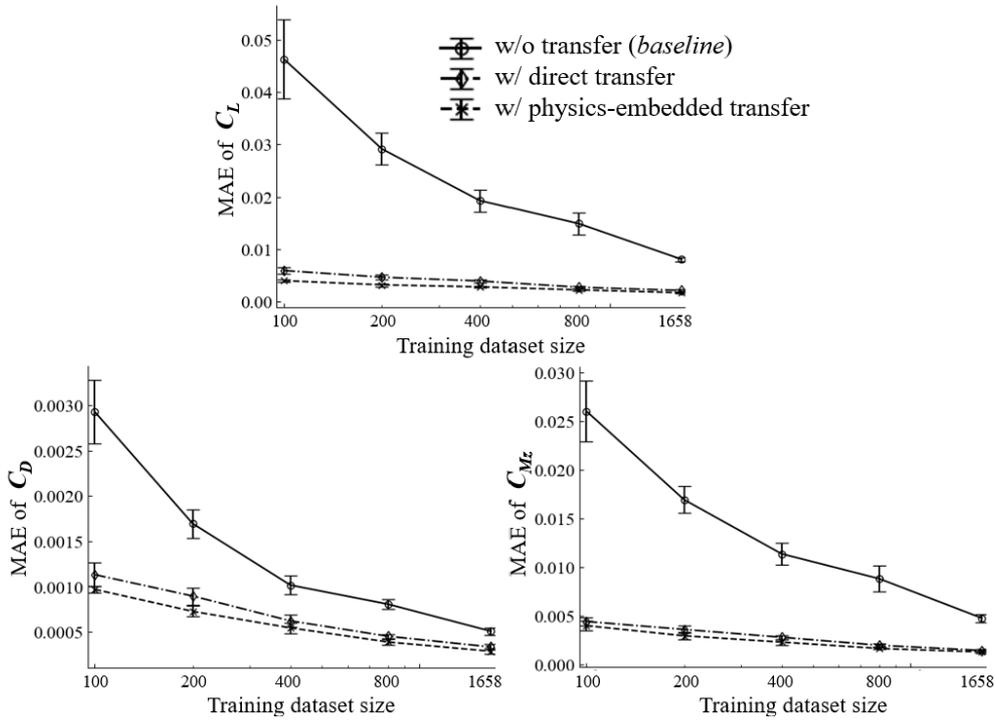

(b) Absolute error of aerodynamic coefficients

**Fig. 12  Influence of training sample size on the prediction errors of frameworks with and without transfer learning**

When reducing the size of the training dataset, the prediction error of the two frameworks that involve transfer learning grows nearly linearly to the log-scaled size for all error metrics. The slopes are smooth, and a relatively small prediction error can be achieved even on the smallest dataset with only 100 wing samples. The difference in the errors between frameworks that do and do not consider single sweep theory does not increase as the dataset size decreases. In contrast, the error of the baseline model increases dramatically when the training dataset size is reduced, and its errors are more than twice that of the transfer framework on small datasets. When comparing the dataset size to achieve a given accuracy, the transfer frameworks can outperform the baseline model in all metrics with less than half of the samples, which proves that the proposed methods can greatly reduce the time cost for establishing enough wing datasets.

## C. Wing aerodynamics prediction with estimated SLD

In the previous section, the model accuracy gains with real SLD input are analyzed. However, in real applications, the SLDs are unknown before prediction, so three ways to estimate the SLD for new wings were proposed in Section II: (1) low-fidelity estimation based on VLM, (2) data-driven estimation based on the auxiliary FNN model, and (3) combination estimation using the FNN model to correct the VLM solution. Their performances are evaluated and compared in this section.

### 1. SLD estimation results

For the SLD estimation method based on the FNN model, the model is also trained during the fine-tuning stage with the same cross-validation groups as the airfoil-to-wing transfer model. For each cross-validation run, the SLD estimation model is trained with the same training wing samples as the airfoil-to-wing transfer model, which guarantees that no information on the testing samples is used during training.

Table 7 shows the SLD estimation errors for different methods. For the low-fidelity VLM estimation, the value in Table 7 indicates the average MSE of SLDs among all wing samples, and for the two model-based methods, the average and standard deviation values are obtained from the test samples among ten cross-validation runs. The SLDs of three randomly selected samples are depicted in Fig. 13, where the SLDs from 3D calculations are shown as black circles, and the SLDs from three estimation methods are shown as colored lines.

Table 7. SLD prediction errors of different estimation methods

|  | low-fidelity | data-driven | combination |
|---|---|---|---|
| Error of SLD | 0.1185 | $0.0106 \pm 0.0003$ | $0.0076 \pm 0.0001$ |

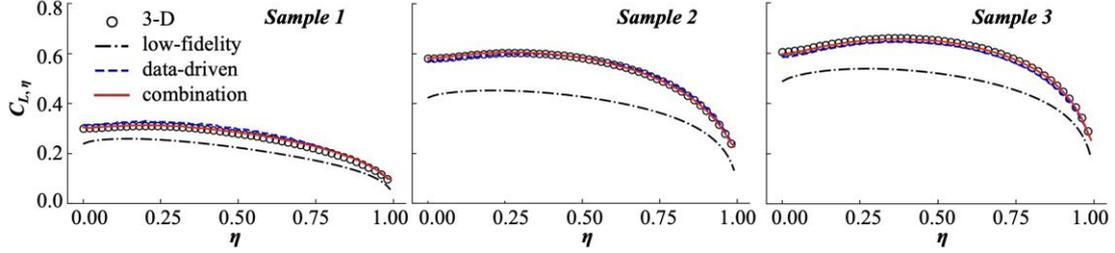

Fig. 13  SLDs estimated with different methods for three test samples

The table and figures indicate that the SLDs from the low-fidelity VLM method have large errors compared to the real 3D ones and always underestimate the lift coefficients because of their hypothesis of a thin wing. The data-driven method provides better estimations, but as shown in Fig. 13, the SLDs estimated with the data-driven method have inconsistent errors spanwise and among different samples. Compared to low-fidelity SLD estimations, this method has random errors without certain patterns, which may cause the airfoil-to-wing transfer model to have difficulty learning the correction, thereby reducing the wing prediction accuracy of the framework. The combination estimation methods combine the advantages of low-fidelity and data-driven methods and can yield the most precise SLD estimates.

## 2. Wing prediction performance comparison for different SLD estimation methods

The three proposed SLD estimation methods are utilized in the physics-embedded transfer learning framework, and the airfoil-to-wing transfer model is trained with the estimated SLDs in the same manner as for the real SLDs in the previous section. Two groups of results from the previous section are selected as the baselines for the evaluation in this section. The nontransfer framework (a) is still used to demonstrate model performance without introducing the transfer learning strategy. The physics-embedded framework (c) with a real 3D simulated SLD is used for reference to show the ideal transfer performance if the SLD is precise. The average and standard deviation of tenfold cross-validation errors are listed in Table 8 in the same manner as in Table 6.

Table 8.    Prediction errors of different SLD estimation methods

| | w/o transfer (baseline) | transferred w/ estimated SLD | | | transferred w/ real SLD (for reference) |
|---|---|---|---|---|---|
| | | low-fidelity | data-driven | combination | |
| $C_p$ | 1.08% ± 0.06% | 0.85% ± 0.06% (-22.1%) | 0.84% ± 0.04% (-23.0%) | 0.77% ± 0.03% (-29.0%) | 0.66% ± 0.04% (-39.3%) |
| $C_{f,\text{sw}}$ | 0.65% ± 0.04% | 0.53% ± 0.04% (-18.3%) | 0.49% ± 0.03% (-24.9%) | 0.46% ± 0.02% (-28.8%) | 0.42% ± 0.02% (-35.6%) |
| $C_{f,z}$ | 0.35% ± 0.02% | 0.30% ± 0.01% (-12.7%) | 0.29% ± 0.02% (-16.4%) | 0.28% ± 0.02% (-20.2%) | 0.26% ± 0.01% (-24.7%) |
| $C_L \times 10^3$ | 8.08 ± 0.37 | 4.94 ± 0.50 (-38.9%) | 6.00 ± 0.58 (-25.7%) | 4.81 ± 0.37 (-40.5%) | 1.70 ± 0.17 (-79.0%) |
| $C_D \times 10^4$ | 5.08 ± 0.41 | 3.55 ± 0.42 (-30.0%) | 3.19 ± 0.37 (-37.2%) | 3.11 ± 0.25 (-38.7%) | 2.87 ± 0.31 (-43.5%) |
| $C_{Mz} \times 10^3$ | 4.77 ± 0.43 | 2.41 ± 0.19 (-49.5%) | 2.66 ± 0.27 (-44.1%) | 2.08 ± 0.19 (-56.4%) | 1.28 ± 0.14 (-73.1%) |

Compared to the transfer learning framework with real SLDs, the framework with estimated SLDs can be applied in practice because it can provide predictions for new wing geometries and operating conditions. For all the performance metrics except for the lift and pitching moment coefficients, the model with the combination SLD estimation method can achieve a similar increase in prediction accuracy as the reference framework with real SLD input. The accuracy gains in the lift and pitching moment are much smaller than those of the reference but similar to those in the drag coefficients. The reason is that the advantage brought from the real SLD input does not exist when the estimated SLDs have errors. The model based on the pure data-driven and low-fidelity (VLM) estimation method achieves a smaller but still considerable improvement in accuracy. The pure data-driven method achieves a larger gain than the low-fidelity method in surface distribution values but generally a smaller gain for integral aerodynamics coefficients.

The three examples used to illustrate model performance with real SLDs in the previous section are used again to show the model performance with estimated SLDs. The model used for comparison here is the transfer framework with the combination SLD estimation method. Fig. 14 shows the surface value contours and cross-sectional value distributions of the three example wings in the same manner as Fig. 11. In subfigure (b), the distributions predicted with the real SLDs are provided in blue lines for reference, and the ones predicted with the combined estimation method are shown in orange lines.

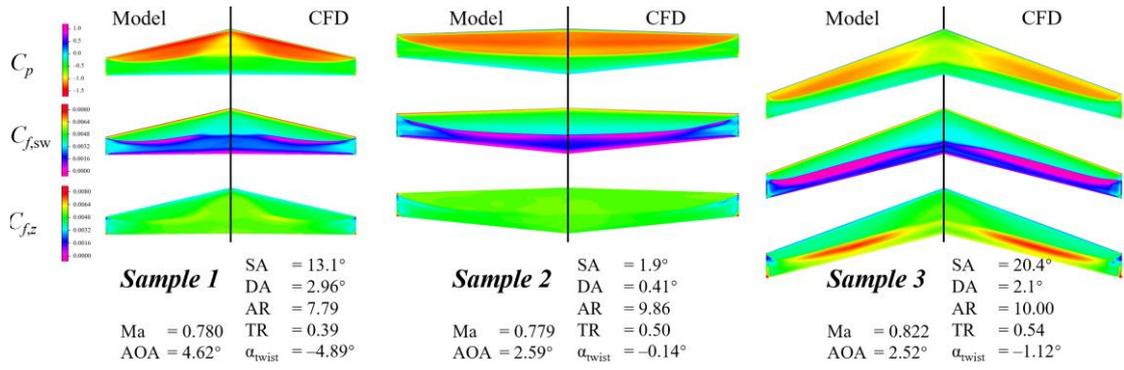

(a) Upper surface contours

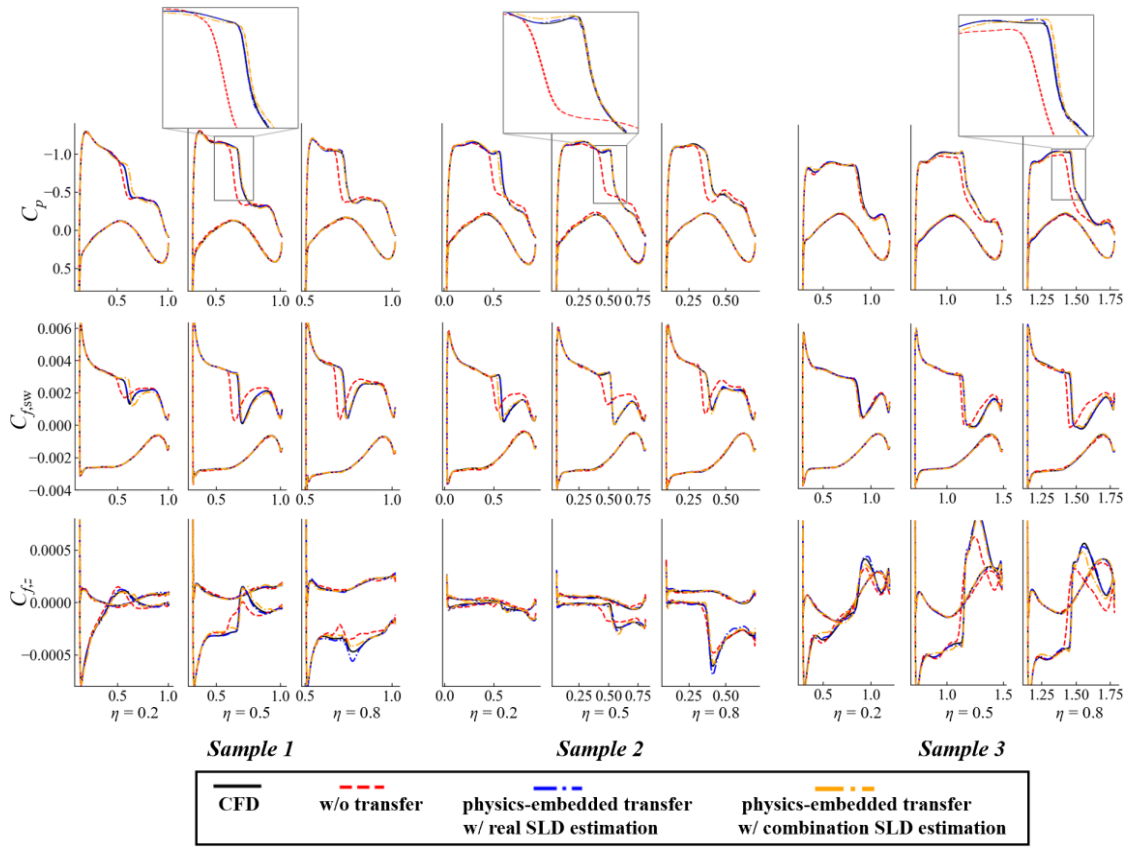

(b) Cross-sectional distributions

**Fig. 14 Results of the example wings predicted with CFD and the transfer model with estimated SLDs**

The results with estimated SLDs are slightly worse in prediction accuracy compared to ones with real SLDs. They have slightly larger errors in key flow features, such as the shock wave intensity and position as shown in the zoom-in plots at the top of Fig. 14, but are still far better than the results from the model without transfer learning. This comparison indicates that the proposed methodology is promising for application in real scenarios.

## 3. Wing prediction performance with different training dataset sizes

The cross-validation scores are plotted against the dataset size in Fig. 15.

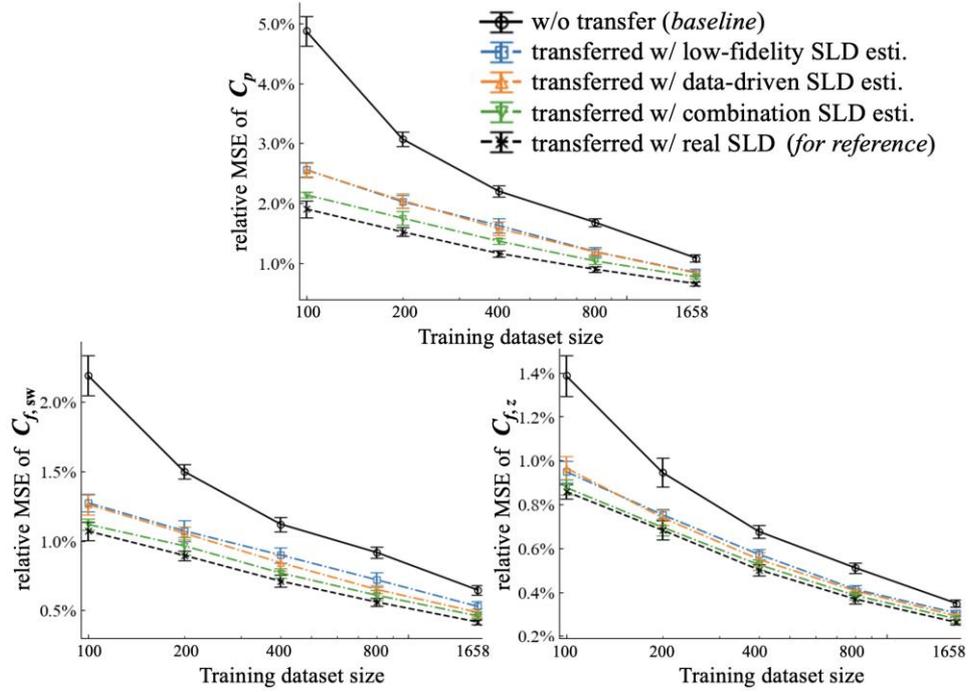

(a) Mean square error of distributed flow quantities

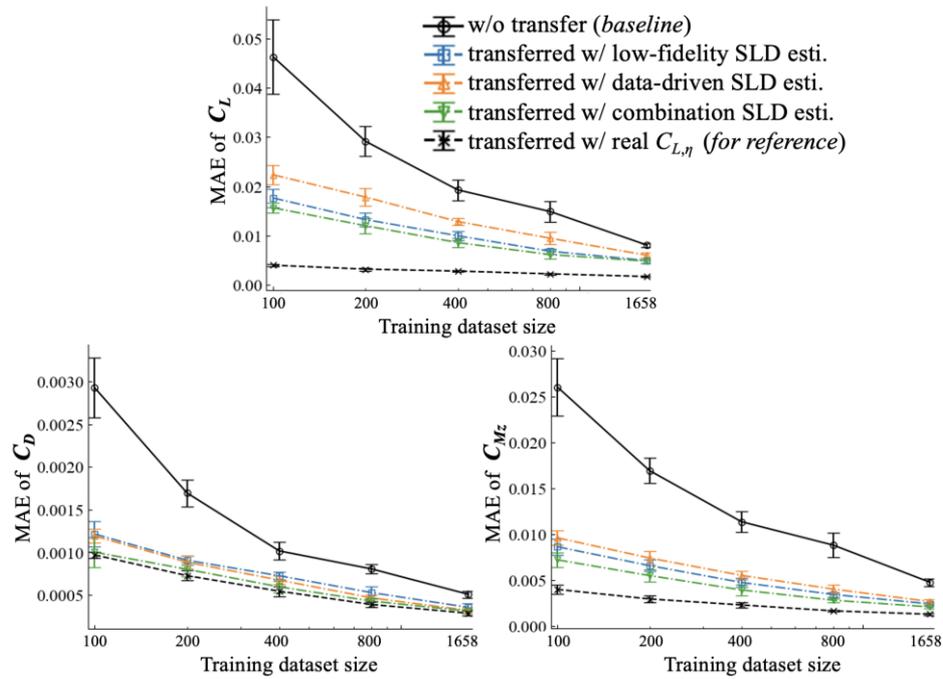

(b) Absolute error of aerodynamic coefficients

**Fig. 15 Influence of training sample size on the prediction errors of frameworks with different lift distribution estimation methods**

As the size of the training dataset decreases, the prediction errors of the models transferred with the three estimation methods still increase linearly with a similar slope as that of the models transferred with real SLDs. Even under the smallest dataset size tested in the present study, all three proposed SLD estimation methods can help the transfer model achieve most of the increase in accuracy. Among the three estimation methods, the combination method yields the best prediction accuracy for almost every variable and dataset size. Its advantage is even greater for small datasets, proving that it is the most promising way to realize physics-embedded transfer learning.

**D. Computational cost**

The computational cost of each step to establish the datasets, train the models, and make predictions are summarized in Table 9. The time costs are listed in the two left columns, and in the four right columns, the required steps by the proposed transfer learning methods and nontransfer methods are ticked.

Compared to the nontransfer framework used to train the ML model for wing aerodynamics prediction, the proposed physics-embedded transfer learning framework requires extra effort to establish the airfoil dataset, obtain low-fidelity SLD results via VLM, and pretrain airfoil aerodynamics models. However, compared to establishing the wing dataset, the computational cost of building the airfoil and SLD datasets can be neglected. In the model training stage, the airfoil model and SLD model only need extra time of less than an hour. Therefore, the reduction brought by the proposed framework in the computational cost of wing samples required to train the model can still bring profits.

For the prediction stage, the machine learning model can make inferences in less than 0.1 seconds for both nontransfer and transfer-learning models. However, the low-fidelity and combination SLD estimation methods require one VLM calculation to get the SLD for wing flow field prediction, which leads to a total prediction time of about 2 seconds. Even so, the machine-learning methods can give almost instant responses compared to the CFD simulation which takes hours. It enables interactive optimization with the model where designers can modify their configurations based on rapid responses of wing flow fields, and it also makes multipoint optimization and robust optimization possible since increasing the design point number will not bring huge computation costs.

Table 9.  The computational cost for different frameworks

| Task | Time cost | w/o transfer | transferred w/ estimated SLD data-driven | low-fidelity | combination |
|---|---|---|---|---|---|
| *Establishing dataset* | | | | | |
| 2-D airfoil dataset (CFD) | 1.2 days [a] | | √ | √ | √ |
| 3-D wing dataset (CFD) | 14.4 days [c] | √ | √ | √ | √ |
| SLD dataset (VLM) | 1.6 hours [b] | | | √ | √ |
| *Training model* | | | | | |
| airfoil aerodynamics model | 0.7 hours [d] | | √ | √ | √ |
| airfoil-to-wing transfer model | 1.7 hours [d] | √ | √ | √ | √ |
| SLD estimation model | 0.1 hours [d] | | √ | | √ |
| *Making prediction* | | | | | |
| CFD simulation (*for reference*) | 15.1 hours (2 procs) | | | | |
| VLM calculation for SLD | 2.2 seconds [b] | | | √ | √ |
| ML model inference | < 0.1 seconds [d] | √ | √ | √ | √ |

a. CFD for airfoils is conducted on Intel® Xeon® Gold 6145 CPU with 32 processors
b. VLM calculation for SLDs is conducted on Intel® Xeon® Gold 6145 CPU with single processor
c. CFD for wings is conducted on Intel® Xeon® Gold 5320 CPU with 160 processors
d. Machine learning model training and inference are conducted on Nvidia RTX 4090 GPU

## V. Conclusions

In the present paper, a physics-embedded transfer learning framework is proposed for predicting the surface pressure and friction distributions of transonic swept wings. Two major ML models are involved in the framework: an airfoil aerodynamics model is first pretrained with 2D airfoil samples to predict the airfoil surface pressure and friction distributions, and then an airfoil-to-wing transfer model is fine-tuned with a few 3D wing samples to predict the wing surface pressure and friction distributions from the results of the airfoil aerodynamics model. During the transfer, the sweep theory is embedded to correct the effective airfoil operating condition that is input to the airfoil aerodynamics model and the output of the airfoil aerodynamics model before being used in the airfoil-to-wing transfer model. Since the spanwise lift distribution (SLD) is needed for sweep theory to correct the values, three methods are proposed to estimate the SLD before the wing surface distributions are predicted. The proposed frameworks are tested via cross-validation, and the conclusions can be summarized as follows:

(1) The wing surface pressure and friction distribution at each spanwise section are determined with local and global (3D) effects. The pretrained aerodynamics model can capture most of the local effects before the airfoil-to-wing model is fine-tuned. The introduction of sweep theory can further capture part of the 3D effects caused by sweep angle. They both reduce the mapping complexity that the model needs to learn, and the testing results show that introducing the pretrained aerodynamics model can reduce prediction errors by 20% to 30%. The introduction of sweep theory can further reduce errors by 4% to 9%.

(2) As the size of the training samples is reduced, the error grows much slower for the physics-embedded transfer learning framework than for the framework without transfer learning. Only less than half of the sample size is required for the transfer learning framework to achieve the same prediction error as the nontransfer learning framework. This approach greatly reduces the need to establish a wing dataset.

(3) Among the three SLD estimation methods, the one that combinates physics-based and data-driven methods achieves the smallest prediction error. It achieves most of the accuracy gain of the framework with real SLD input on small and relatively large datasets. The proposed SLD estimation method enables the application of the physics-embedded framework in practice.

In summary, the proposed physics-embedded transfer learning framework leverages the traditional knowledge of the relationship between the airfoil and wing flow fields and reduces the need for samples to train a flow field prediction model for wings. However, only the single-segment wings are studied in the present paper where the gap still exists between them with realistic wing configurations with kink and interaction with other aerodynamic components. In Ref. [31], the authors proposed another transfer learning strategy, which can be combined with the one proposed in this paper to efficiently expand the wing design space to kink wings, and for more complex configurations, we believe the same idea of 2D-to-3D transfer learning still works. On the other hand, the network architecture used in the present study is rather simple, and the hyperparameters are tuned manually. By further optimizing the network architecture, better prediction errors may be obtained, and it will be possible to apply the physics-embedded transfer learning framework in downstream tasks such as aerodynamic shape optimization.

## Acknowledgments

This work was supported by the National Natural Science Foundation of China, Nos. 92052203, 12202243, 12372288, 12388101, and U23A2069.

## Data Availability

The data that support the findings of this study are available from the corresponding author upon reasonable request.

# Appendix A. Architectures of the machine learning models

The architectures for the models involved in the physics-embedded transfer learning framework are introduced here. All the codes are implemented with PyTorch and are open-source to the GitHub repository *flowGen*.

### 1. Airfoil aerodynamic model

The prediction of the aerodynamics of 2D airfoils is fulfilled with a U-Net framework[34]. The input of the model is the airfoil geometry, which is described by the surface grid points, as well as the operating conditions. The output is the pressure and friction distribution on the same surface grid points as the input. Thus, the input and output are one-dimensional (1D) data.

The framework of the model is illustrated in Fig. A1 and comprises an encoder and a decoder composed of 1D convolution blocks with a symmetric architecture. The encoder contracts the airfoil geometry to the low-dimensional representation $z_f$, which is concatenated with the operating conditions and then input to the decoder to generate the target pressure and friction distribution. Skip connections are established between the encoder and decoder as the major feature of the U-Net framework. The feature maps along the contracting path in the encoder are saved and reinjected into the upsampling path in the decoder by concatenating them to the symmetric level. Several previous studies [35]-[36] have shown that the U-Net framework can improve model performance.

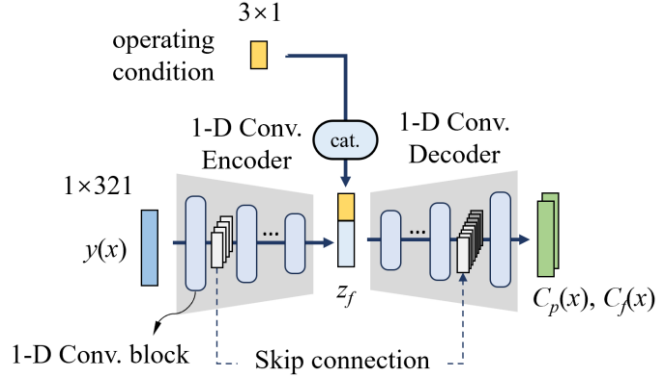

Fig. A1　Framework of the airfoil aerodynamics model

The encoder comprises three blocks, each consisting of a 1D convolution layer with a kernel size ($k$) = 3 and a stride ($s$) = 2, an averaged pooling layer with the same kernel size and stride, and a leaky ReLU activation layer. The channel numbers after each block are 16, 32, and 64, and the feature map sizes are 80, 19 and 4. A densely connected layer then links the flattened output of the last encoder block to a 32-dimensional latent variable $z$.

The decoder shares a similar architecture as the encoder, where an upsampling layer replaces the 1D convolution layer. The upsampling layer consists of one linear interpolation layer for upsampling the 1D data and a convolution layer with $k = 3$ and $s = 1$ to manufacture features. The channel numbers are 512, 256, 128, and 64, and the feature map sizes are 4, 19, 80, and 321. Before the first block, two densely connected layers with 64 and 128 hidden dimensions are used to connect the latent variable and the initial feature map. After the last block, another convolution layer with $k = 3$ and $s = 1$ compresses the last feature map to two channels.

The channel numbers and feature map dimensions of the encoder and decoder are tuned manually with several trial-and-error experiments to achieve the best performance on the validation samples.

2.　Airfoil-to-wing transfer model

The airfoil-to-wing transfer model is also realized with the U-Net framework in the present paper and is illustrated in Fig. A2. The hyperparameters in the model are also tuned manually with several trial-and-error experiments.

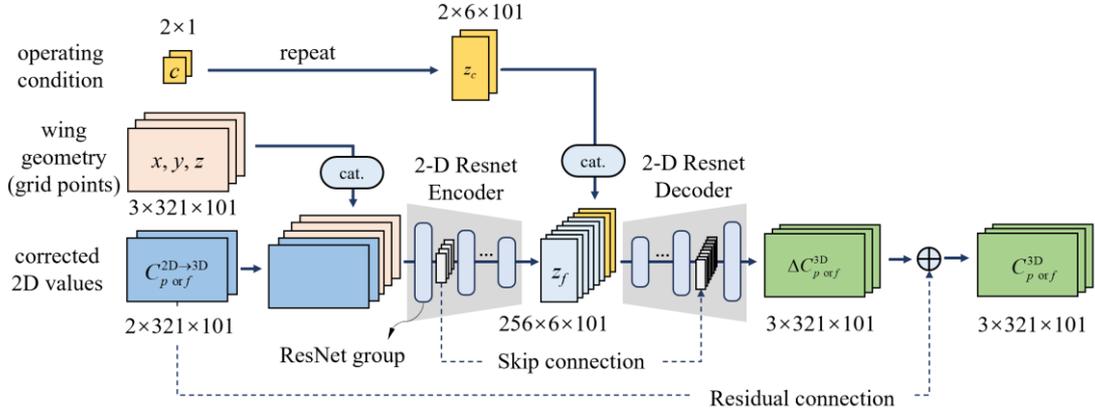

Fig. A2　　Framework of the airfoil-to-wing transfer model

The encoder and decoder manipulate only the airfoil circumferential dimension in feature maps, and the spanwise dimension remains the same. Thus, the latent variable in the present framework is no longer in vector form but rather a compressed feature map with three dimensions. This approach saves the model from having numerous parameters caused by the fully connected layers between compressed feature maps and the latent vector and helps the model better preserve the information from airfoil aerodynamic model results on each spanwise location.

The wing geometry is described as the coordinate of its surface mesh points and concatenated with the corrected 2D distributions before being input to the encoder. The surface mesh points are used for two reasons. First, the mesh has the same size as the corrected 2D distributions, which makes the effect of introducing 2D distributions easy to evaluate. The baseline model can be constructed by simply eliminating corrected 2D distributions from the inputs, and the model backbone mostly remains the same, making it a fair comparison. Second, in our previous study [31], it was proven that a distributed geometric input is better than only inputting the overall geometric parameters. The surface mesh points can be considered the distributed geometric input, which helps the model achieve better accuracy.

The operating conditions ($Ma^{3D}$, $AOA^{3D}$) are input to the model by concatenating to the latent variables inside the U-Net. Since the latent variables in the present framework are a compressed feature map, the operating conditions are repeated airfoil-circumferential-wise and spanwise to obtain the same size as the latent feature maps.

Since the model has a larger size and dimension, 2D ResNet blocks[37] are used to build the encoder and decoder of U-Net to mitigate the training difficulty caused by the increasing number of layers. The encoder and decoder are composed of six ResNet groups. Each ResNet group in the encoder contains a residual block for downsampling and a basic residual block, whose architectures are illustrated in Fig. A3. The size in the airfoil-circumferential dimension is halved by each group, and the channel numbers after the six groups are 32, 64, 64, 128, 128, and 256. In the decoder, the first residual block in each ResNet group is replaced by a residual block for upsampling realized with convolutional layers and linear interpolation, which is also shown in Fig. A3. In contrast to the encoder, the feature map size is doubled by each group in the decoder, and the channel numbers after the six groups are 128, 128, 64, 64, 32, and 32. The decoder finalizes with a single convolutional layer that compresses the 32 channels to three corresponding to the output variables.

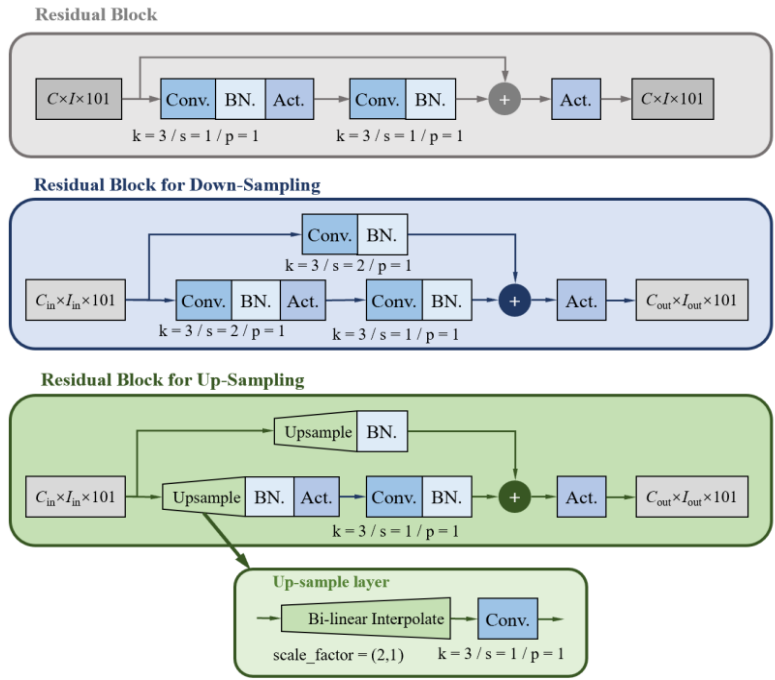

**Fig. A3   Architectures of residual blocks for the airfoil-to-wing transfer model**

### 3.  Spanwise lift distribution model

The two data-driven methods are proposed to estimate SLDs for building the relationship between the 2-D and 3-D results: one is to directly predict SLDs, and the other is to predict the residual between real SLDs and the VLM-predicted ones. The two methods share the same machine learning model architecture to predict the results. It is designed to be simpler than the former two models because the SLD has a simpler pattern than the surface pressure and friction distributions and because of the limited size of the training samples. As shown in Fig. A4, a multi-input feedforward neural network (FNN), whose architecture is shown in Fig. A4, is used. The input is divided into three parts: the operating conditions, including $Ma^{3D}$ and $AOA^{3D}$; the wing planform geometry, including $\Lambda_{LE}$, $\Gamma_{LE}$, AR, TR, $r_t$, and $\alpha_{twist}$; and the sectional airfoil geometry, including the CST parameters and $(t/c)_{max}$ for the root. The three parts of the input are passed through three separate densely connected layers with hidden dimensions of (16), (32, 32), and (64, 64) before they are concatenated and passed through two densely connected layers with a hidden dimension of 64 and an output dimension of 101. For the proposed method (2), the output is set to be the SLD predicted by CFD simulations, and for method (3), the output is the difference between the VLM and CFD results.

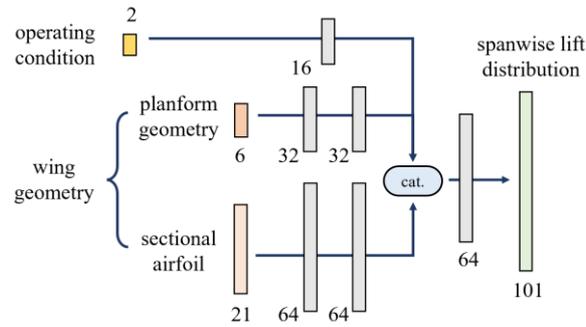

**Fig. A4    Framework of the lift distribution prediction model**